\documentclass[pre,aps,twocolumn,showpacs]{revtex4-1}
\usepackage{graphicx,epsfig,amsmath,amsfonts,bm}

\def\text#1{{\mathrm{#1}}}

\begin{document}

\title{A simple deterministic and time reversal invariant thermostat}
\author{Henk van Beijeren}
\email{H.vanBeijeren@uu.nl}
\affiliation{Institute for Theoretical Physics, Utrecht University, Leuvenlaan 4, 3584 CE, Utrecht, The Netherlands}
\begin{abstract}
\noindent
A simple deterministic and time reversal invariant type of thermostat is proposed to be used for computer simulations of classical systems. It acts on collisions with the walls of the container exclusively. It maps the incoming and outgoing velocity of an impinging particle onto each other uniquely, in a way that satisfies a detailed balance condition with respect to the (local) wall temperature. It is fast to implement on a computer, leads to rapid equilibration or approach to a stationary nonequilibrium state and it effects the physical properties of the system in a narrow boundary layer only. Because of the deterministic nature it is especially suitable for studying dynamical systems characteristics, such as Lyapunov exponents and fractal dimensions of attractors. A few successful applications have been reported already.
\end{abstract}

\date{\today}

\maketitle

In computer simulations of classical many-particle systems the most common way of accounting for energy and momentum exchange with the walls of the system has been by modelling particle-wall collisions stochastically (see e.g.\cite{vBD}). Usually, in these models a particle colliding with a wall is supposed to be absorbed and reemitted instantaneously according to a Maxwell distribution with the local temperature and velocity of the wall. This makes the implementation in a simulation rather slow; for each velocity component a random number has to be drawn and converted to a velocity value. And in applications to dynamical systems, where one likes to follow bundles of nearby trajectories in phase space it is not clear how to implement  the random reflections in a consistent way to a full bundle.

An alternative exists in the form of deterministic thermostats, modelling collisions of particles with a wall by mapping precollisional velocities in a unique way to postcollisional ones (see e.g.\cite{Rondoni}).
For example, Chernov and Lebowitz\cite{CL} studied Couette flow by introducing deterministic wall thermostats, which assign to each incoming velocity $\bm{v}$ a unique outgoing velocity $\bm{v}'$, depending on the orientation and local velocity of the wall at the position $\bm{q}$ of the collision. This is done in one case in a time reversal invariant way, so an incoming velocity  $-\bm{v}'$ is mapped to $-\bm{v}$. In addition $|\bm{v}|=|\bm{v}'|$, so energy is conserved and the mapping is such that the mean velocity of a particle at the wall is proportional to the local wall velocity. In case the latter vanishes uniformly, the microcanonical distribution (if necessary constrained to zero momentum parallel to the walls and fixed value of the corresponding center of mass component) is stationary under the equations of motion, and initial nonequilibrium distributions should decay to equilibrium (in the usual coarse-grained sense). Under Couette boundary conditions rapid decay to a macroscopically stationary state has been observed, accompanied by an ongoing phase space contraction toward a fractal distribution of lower dimensionality than that of phase space\cite{CL,PD}.

For simulating thermal walls, which exchange energy with the system in such a way that at least locally it tends to equilibrate at a temperature $T(\bm{q})$, similar mappings may be devised. The simplest of these probably is a mapping in which the normal velocity, $v_n=\bm{v}\cdot\hat{\bm n}(\bm{q})$, with $\hat{\bm n}(\bm{q})$ the unit vector normal to the wall pointing inward at the position $\bm{q}$, is mapped to $v'_n$, and the velocity components parallel to the wall remain unchanged. This mapping is defined for $v_n<0$ and has to result in a positive $v'_n$. Time reversal invariance is obtained under the condition $(-v'_n)'=-v_n$. For a wall at uniform temperature the canonical distribution ought to be
stationary under the wall collisions (together with the Hamiltonian dynamics in the interior, which indeed leave the canonical distribution invariant). This can be obtained by imposing  a detailed balance condition, requiring that in equilibrium the frequencies of collisions with impact velocities $\bm{v}$ and $-\bm{v}'$ are equal. Then the number of particles coming off the wall with velocity $\bm{v}'$ in equilibrium is the same as under e.g.\ specular reflections.
 The average number of collisions per unit wall area with velocity between $\bm{v}$ and $\bm{v}+d\bm{v}$  is given in general by
\begin{equation}
n_c(\bm{v},\bm{q})d\bm{v}=-v_nf^{(1)}(\bm{v},\bm{q})d\bm{v},
\label{one}
\end{equation}
with $f^{(1)}(\bm{v},\bm{q})$ the one particle distribution function. In the canonical ensemble the latter is of the form
\begin{equation}
f^{(1)}_{can}(\bm{v},\bm{q})=n(\bm{q})\left(\frac{\beta m}{2\pi}\right)^{d/2}\exp\left(-\frac{\beta m v^2}2\right),
\label{}
\end{equation}
with $d$ the dimensionality of the system, $\beta=1/(k_B T)$ and $k_B$ Boltzmann's constant. Hence one may express the detailed balance condition as
\begin{equation}
-v_n\exp\left(-\frac{\beta m v_n^2}2\right) |dv_n|=v'_n\exp\left(-\frac{\beta m(v'_n)^2}2\right) |dv'_n|.
\label{three}
\end{equation}
The simplest nontrivial  solution of this, mapping the interval $(-\infty,0]$ to $[0,\infty)$ is of the form\cite{footnote1},
\begin{equation}
\exp\left(-\frac{\beta m( v'_n)^2}2\right) =1-\exp\left(-\frac{\beta m v_n^2}2\right) ,
\label{four}
\end{equation}
or,
\begin{equation}
v'_n=\sqrt{-\frac2{\beta m}\ln\left(1-\exp\left(-\frac{\beta m v_n^2}2\right)\right)}.
\label{five}
\end{equation}
In case the wall temperature is not uniform one simply has to replace $\beta$ by $\beta(\bm{q})$ in this expression. Notice that this mapping satisfies time reversal invariance.

This thermostat, and variations of it have already been used a few times. Garrido et al\cite{Garrido} employed it in a simulation of heat conductivity in one-dimensional chains and Bosetti and Posch used a variant in a study of heat conduction and the corresponding dynamical system properties of hard disk systems\cite{Bosetti}. In this variant the collisions remain specular in direction, but the speed is changed in such a way that the desired detailed  balance condition is satisfied.  In both cases the authors observed a rapid approach of the system to the stationary state. From Eq.\ (\ref{four}) one easily sees that the thermostat converts distributions of the normal velocity with average energy $>k_B T/2$ to distributions with average energy $<k_B T/2$ and vice versa. This, by itself unphysical effect no doubt contributes to the rapid convergence. And outside the narrow boundary layers it will not affect the temperature profile nor other system properties.

In principle the thermostat described here could also be used for solid or glassy systems by confining these between rigid walls. However, the restriction to classical equations of motion will be problematic for many of these systems. 

The thermostat also offers an alternative to Nos\'e-Hoover thermostats for sampling the canonical ensemble with deterministic dynamics, provided the system to be studied can be designed to have a hard wall somewhere.

In conclusion, this paper describes a deterministic thermostat that is very efficient in equilibrating classical systems near thermal walls, that can be simulated fast on a computer and is especially useful for investigating dynamical system properties.


\begin{thebibliography}{9}
\bibitem{footnote1} There are other solutions with an alternation of intervals with $v'_n=-v_n$ and intervals with\\ $v'_n=\sqrt{-\frac2{\beta m}\ln\left(1-\exp\left(-\frac{\beta m v_n^2}2\right)\right)}$. \
\end{thebibliography}
\end{document}